\documentstyle[prl,twocolumn,aps,epsf]{revtex}

\begin{document}

\twocolumn[\hsize\textwidth\columnwidth\hsize\csname @twocolumnfalse\endcsname
\draft
\tolerance 500

\title{Quantum dynamics in two and three-dimensional quasiperiodic tilings}
\author{Fran\c cois Triozon$^{1}$, Julien Vidal$^{2}$, R\'emy
Mosseri$^{2}$ and Didier
Mayou$^{1}$}

\address{
$^{1}$ LEPES-CNRS, 25 avenue des Martyrs BP166, 38042 Grenoble, France \\
$^{2}$ Groupe de Physique des Solides, CNRS UMR 7588, Universit\'{e}s
Paris 6 et  Paris 7,\\
2, place Jussieu, 75251 Paris Cedex 05 France}

\maketitle

\begin{abstract}

We investigate the properties of electronic states in two and three-dimensional
quasiperiodic structures~: the generalized Rauzy tilings. Exact diagonalizations,
limited to clusters with a few thousands sites, suggest that eigenstates are
critical and more extended at the band edges than at the band center. These
trends are clearly confirmed when we compute the spreading of energy-filtered
wavepackets, using a new algorithm which allows to treat systems of about one
million sites. The present approach to quantum dynamics, which gives also access
to the low frequency conductivity, opens new perspectives in the analyzis of two
and three-dimensional models.

\end{abstract}


\vskip2pc]
%
%
%
%

Since the discovery of quasicrystals \cite{Shechtman}, the nature of the eigenstates
and of the energy spectrum, as  well as the propagation of electrons in
quasiperiodic potentials, have been  the subject of many theoretical studies.
One-dimensional ($1D$) quasiperiodic Hamiltonians have been widely investigated by
means of numerical \cite{Kohmoto83,Ostlund} and perturbative approaches
\cite{Sire_pertu,Piechon}.  Their spectrum consists in a set of zero width bands
associated to critical eigenstates responsible of a sub-ballistic propagation of the
electrons. In higher dimensions, the situation is more complex and, given the
geometrical complexity of the structures, analytical treatments are difficult. Most
of the studies have focussed on topologically trivial systems whose properties can
be easily extracted from the $1D$ case (Fibonacci quasilattices \cite{Ueda,Zhong},
labyrinth tiling
\cite{Sire_Lab,Yuan}). It has been shown that if the propagation is always
sub-ballistic, the spectrum could be either absolutely continuous, singular
continuous or any mixture. Other works based on exact diagonalizations, have also
been performed on Penrose-like $2D$ and $3D$ systems. In $2D$, there is some
evidence that eigenstates are critical
\cite{Yamamoto_Penrose} and responsible of an anomalous diffusion
\cite{Passaro_Octo}.  In $3D$ systems, the nature of eigenstates remains unclear
\cite{Rieth_Penrose2D3D}. In this context, it is of great interest to determine
whether critical states are generic of {\em true} quasiperiodic systems especially
in the $3D$ case, and to study the physical properties induced by such states.

In this letter we study in a tight-binding approach, the electronic properties of the
of the $2D$ and $3D$ gene\-ralized Rauzy tilings (GRT) \cite{Vidal_Rauzy}. We discuss
the nature of the eigenstates by computing their localization degree for small
systems ($\sim 5500$ sites). This analy\-zis suggests that the eigenstates are
critical and rather more extended at the band edges than at the band center. To go
beyond these exact diagonalizations, we introduce a powerful algorithm that allows
us to study the quantum dyna\-mics for very large systems (up to
$10^{6}$ sites here). Sub-ballistic propagation laws are clearly established and the
energy dependence of the exponents confirms that states are more mobile (almost
ballistic) close to the band edges than in the center. From this point of view,
these results provides evidence of critical states in the $2D$ and $3D$ GRT.
Finally, we discuss the low frequency conductivity which can be directly related to
the quantum diffusion \cite{Sire_Aussois,Bellissard_anomalous,Mayou00}.

%
%
%
%

\begin{figure}
\centerline{\epsfxsize=58mm
\epsffile{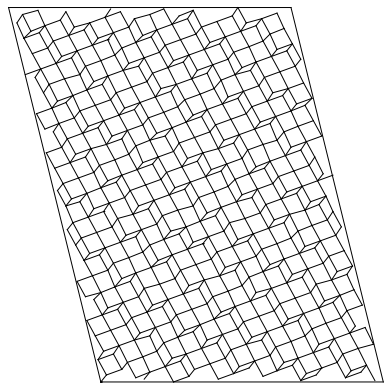}}
\caption{A unit cell of the $10^{\rm th}$ approximant of the $2D$ GRT
with sites.}
\label{fig:2Dtiling}
\end{figure}

The GRT are codimension one quasiperiodic structures that can easily be
built in any dimension $D$ by the standard cut and project method
\cite{Vidal_Rauzy}. Here, we focus
on approximant structures that are obtained for rational cut
directions. These tilings have a complex topological structure with
sites of coordination
number ranging from $D+1$ to $2D+1$.
as displayed in Fig.~\ref{fig:2Dtiling} for $D=2$.
%
%
%
%
This geometrical complexity implies that they cannot be
separated into a combination of $1D$ systems as in the  Fibonacci
quasilattices \cite{Ueda} or
as in the labyrinth tiling \cite{Sire_Lab}. We thus expect that their
spectral and dynamical
properties are representative of other $2D$ or $3D$ Penrose-like quasiperiodic
structures.
%
%
%
%

\begin{figure}
\vspace{-3mm}
\centerline{\epsfxsize=100mm
\hspace{50mm}
\epsffile{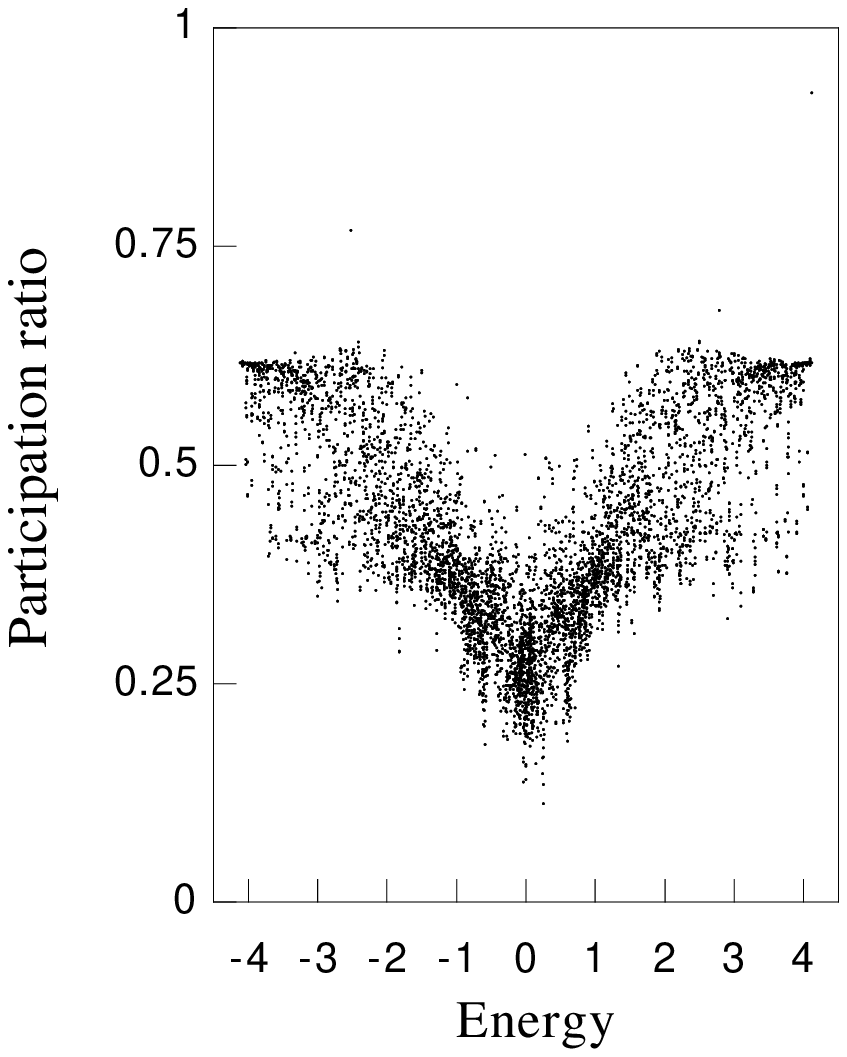}
\epsfxsize=100mm
\hspace{-156pt}
\epsffile{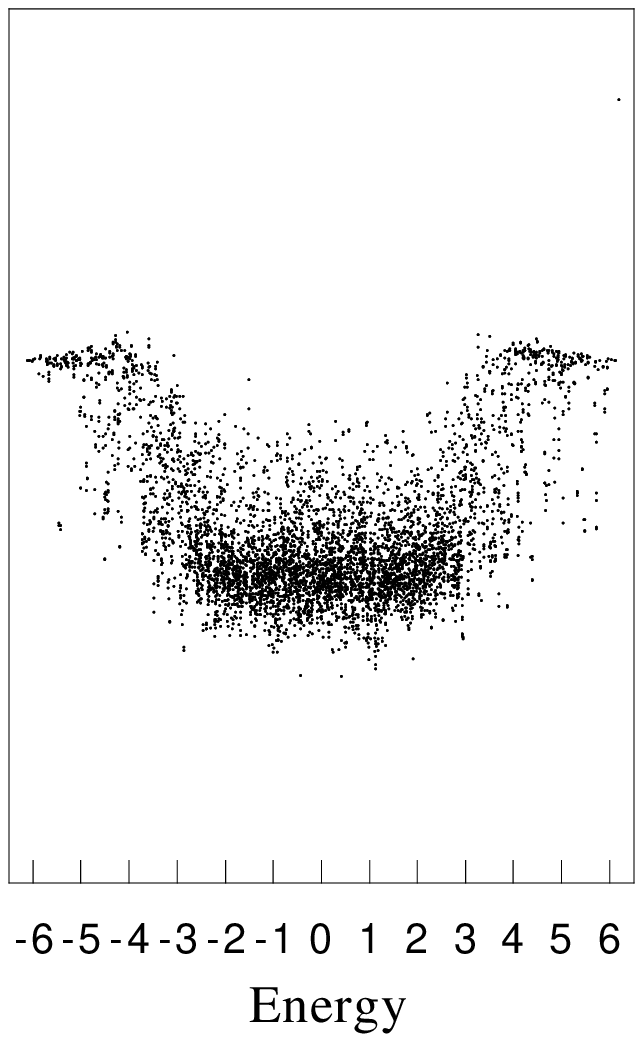}}
\vspace{-6mm}
\caption{Participation ratio as a function of the energy for the
$14^{\rm th}$ approximant (5768
sites) of the $2D$ GRT (left) and for the $13^{\rm th}$ approximant (5536
sites) of the $3D$ GRT
(right).}
\label{P(E)}
\end{figure}

We consider noninteracting electrons described by the following
tight-binding Hamiltonian~:
%
%
\begin{equation}
H = \sum_{\langle  i,j\rangle} t\, |i \rangle \langle  j|
\mbox{,}
\end{equation}
%
%
where $t=1$ is the hopping energy between nearest neighbor sites, and where $|i
\rangle$ denotes an orbital localized on the site $i$.  As discussed in
\cite{Vidal_Rauzy}, the Hamiltonian matrix with periodic boundary conditions is
straightforwardly written as a band (Toeplitz-like) matrix. This is
due to a natural
indexing of the sites known as the conumbering which classify them
according to their
local environment. The study of the spectrum of $H$ has already been
presented in
Ref.~\cite{Vidal_ICQ7}, and we focus here on the nature of the eigenstates.
Therefore, we have computed by exact numerical diagonalization, the
participation
number $P$ of each normalized eigenstate
$|\psi\rangle$ defined by~:
$P(\psi)=\left(\sum_i |\langle  i| \psi \rangle|^4 \right)^{-1}$. For
a given system
size, this quantity measures the localization degree of the state
$|\psi\rangle$ considered. We display in Fig.~(\ref{P(E)}), for the
$2D$ and $3D$ GRT, the participation ratio $p=P/N$ as a function of
the energy ($N$
is the number of sites per unit cell). Qua\-litatively, we observe in
both systems
that, despite strong fluctuations, the most extended eigenstates are
globally located
near the spectrum edges and the more localized ones in the central part.
%
%
%
%

\begin{figure}[h]
\centerline{\epsfxsize=95mm
\hspace{5mm} \epsffile{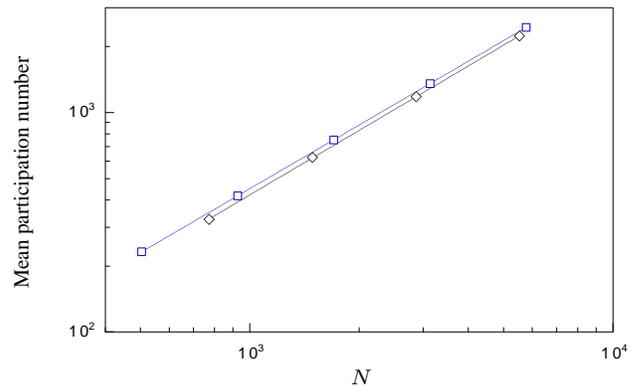}}
\vspace{-8mm}
\caption{Mean participation number as a function of the number of
sites for the $2D$ ($\Box$)
and the $3D$ ($\Diamond$) GRT.}
\label{P(N)}
\end{figure}

Another interesting information is given by the sca\-ling of the mean
participation number (averaged
over the whole spectrum) versus the number of sites $N$. Indeed,
when the system size increases, the
mean participation number converges towards a finite value if the states
are exponentially localized and goes to infinity otherwise.  As
shown in Fig.~\ref{P(N)}, this mean partici\-pation number
behaves as $N^{\gamma}$ with $\gamma_{2D} \simeq 0.965 \pm 0.001$ and
$\gamma_{3D}=0.975 \pm 0.003$, indicating a
weakly critical behaviour of the eigenstates.  Note that this is only
an average behaviour that conceals strong
fluctuations from one energy to another as it has already been shown
in Penrose-like tilings \cite{Rieth_Penrose2D3D}.

In the $2D$ GRT, the exponent is greater than the one obtained for
the octagonal tiling
\cite{Grimm_ICQ7} (codimension 2) where $\gamma=0.87 \pm 0.05$.
This difference can be understood by invoking the codimension of the
structures that gives an indication on the geometrical complexity in
terms of local environment.
Intuitively, we indeed expect low codimension structures to be weakly
{\em quasiperiodic}, and to have weakly {\em critical} states. The recent level
statistics analyzis done in the $2D$ GRT \cite{Jagannathan_Rauzy}
also corro\-borates this
assumption. Finally, up to the numerical precision, we have
$\gamma_{2D}< \gamma_{3D}$. It
means that for a fixed codimension (here 1), the localization of the
eigenstates increases
when the dimension is lowered, as it is the case in disordered systems.
To confirm the trends observed with diagonalizations, we have studied the
spreading of energy-filtered states on much larger systems of about
one million sites.
%
%
%
%

The diffusion of wave packets can be measured by the mean square spreading~:
%
%
\begin{eqnarray}
X^{2}(E,t) &=& \langle (\hat X(t)-\hat X(0))^{2}\rangle_{E}
\label{eq:spread1}\\
&=&\langle A^{\dagger}(t)A(t)\rangle_{E}
\mbox{,}
\label{eq:def2}
\end{eqnarray}
%
%
where $A(t)=[\hat X,\mbox{e}^{-\mbox{i}Ht}]$. $\hat X(t)$ is the Heisenberg
representation of the position operator along the $x$ direction
and  $\langle\dots \rangle_{E}$ denotes the average value over
the eigenstates of energy $E$. We stress that $X^{2}(E,t)$ evolves
with time even though it is a mean value over eigenstates.

$X^{2}(E,t)$ is related to $\sigma(E_{\rm F},\omega)$ the conductivity at
frequency $\omega$ and Fermi energy $E_{\rm F}$ \cite{Mayou00}. In the limit of
small frequencies one has~:
%
%
\begin{equation}
\langle \sigma (E_{\rm F},\omega)\rangle_{\Delta} =
e^{2}\int\limits_{0}^{\infty} \mbox{e}^{\mbox{i}\omega t} \left\langle
n(E_{\rm F}){\mbox{d}^{2}X^{2}(E_{\rm F},t)\over
\mbox{d}t^{2}}\right\rangle_{\Delta} {\rm d}t
\label{eq:COND}
\end{equation}
%
%
where $n(E_{\rm F})$ is the density of states at the Fermi energy, $e$ is the
electron charge, and $\langle \dots \rangle_{\Delta}$ means an average over a
range $\Delta $ of values of $E_{\rm F}$. Note that at zero frequency,
the integral in
(\ref{eq:COND}) is equal to the diffusivity
$ D(E_{\rm F})= \mbox{lim}_{t \to \infty} \mbox{d} X^{2}(E_{\rm F},t)/ \mbox{d} t$
and one recovers the Einstein formula  $\sigma (E_{\rm F},0)= e^{2}
n(E_{\rm F}) D(E_{\rm F})$.

To investigate the quantum diffusion, we consider an approximant of
the GRT with a
sufficiently large unit cell so that the eigenstates are Bloch waves
characterized by a wave
vector ${\bf k}$. As for the exact diagonalizations we  restrict our
study to the ${\bf k} = 0$ subspace, i.e. to the eigenstates which
are invariant by
translation from one unit cell to another. Note that for an approximant the
operator $A(t)$ is invariant by translation of a unit cell and thus
lets invariant the
subspace of ${\bf k} = 0$ states. We calculate $X_{0}^{2}(E,t)$ defined by~:
%
\begin{eqnarray}
X_{0}^{2}(E,t)&=&\langle A^{\dagger}(t)A(t)\rangle_{E,{\bf k} = 0}
\label{eq: def4}\\
X_{0}^{2}(E,t)&=& {\mbox{Tr} \left\{A^{\dagger}(t)\delta(E-H)A(t)\right\}
\over
\mbox{Tr} \left\{\delta(E-H) \right\}}
\label{eq: def33}
\end{eqnarray}
%

Actually, $X_{0}^{2}(E,t)$ is a good approximation of  $X^{2}(E,t)$
provided that $t$ is lower than the time required to significantly
couple orbitals
separated by a distance $l_x/2$ during the time evolution. In this
study, we have
always considered wave packet evolutions for such times.
The operator trace $\mbox{Tr}$ is restricted to the subspace ${\bf k} = 0$. An
efficient trick to evaluate the traces is to use states $|\psi\rangle$
having a constant presence probability on each site $|\langle i |
\psi \rangle|^2=1/N$ and a random
phase on each orbital of the unit cell. We thus
approximate
$\mbox{Tr}\{ B \}$ by $N \langle \psi | B| \psi\rangle$. In fact, we
should perform
an average over several random phases configurations but we have checked that,
thanks to the great number of
sites in the unit cell, we only need to consider one random phase
state $|\psi\rangle $ to get
a good estimate of the traces.  Moreover, to eliminate
fluctuations, we have performed a
convolution of the $\delta (E-H)$ by a Gaussian operator of typical
width $\Delta=W/100$ where $W$ is the total
bandwidth \cite{Triozon_RIKEN}. This is equivalent to compute the spreading
of energy filtered wavepackets. These parts of the calculation are rather
standard and in fact the crucial step is to evaluate $A(t)|\psi\rangle$.

We start from the expansion of the
evolution operator $\mbox{e}^{-\mbox{i}Ht}$ using a
basis of Chebyshev polynomial $P_{i}(E)$ of the first kind. These
polynomials are associated to the weight 
$\rho(E) = 1/\left( \pi \sqrt {4b^{2}-(E-a)^{2}}\right)$. This expansion converges
provided the spectrum of $H$
(computed here using the standard recursion method
\cite{Haydock_rec3}) is included in
the interval $[a-2b,a+2b]$ \cite{Triozon_RIKEN}.
The decomposition of the evolution operator is written as~:
%
%
\begin{equation}
\mbox{e}^{-\mbox{i}Ht}=\sum_{i=0}^{\infty} c_{i} (t) P_{i}(H)
\mbox{,}
\label{eq: expU}
\end{equation}
%
%
with $c_{p}(t)=\mbox{i}^{p} h_{p} J_{p}(-2bt)$ where $h_{0}=1$, $h_{p>0}=1/2$,
and where $J_{p}(x)$ is the
Bessel function of order $p$ \cite{Numerique,Roche_Fibo_desordre}. At
large $p$ and fixed
$x$, one has~:
$J_{p}(x)\simeq 1/\sqrt{2\pi p}\, (x\, \mbox{e}/2p)^{p}$ so that the sum in
Eq.~(\ref{eq: expU}) converges quickly and can be truncated.
The polynomials $P_{i}$ obey the following relation~:
%
%
\begin{equation}
b P_{i+1}(H) = (H-a) P_{i}(H)- b P_{i-1}(H)
\label{eq: RecPo}
\mbox{,}
\end{equation}
%
%
with $P_{0}(H)=1$ and $P_{1}(H)=(H-a)/2b$.
Then, introducing $|\phi_{i}\rangle = P_{i}(H) \, |\psi\rangle$ and
$|\psi_{i}\rangle =[\hat X,P_{i}(H)]\, |\psi\rangle$ and using
Eqs.~(\ref{eq: expU}) and
(\ref{eq: RecPo}), one readily obtains~:
%
%
\begin{eqnarray}
b|\phi_{i+1}\rangle &=& (H-a)|\phi_{i}\rangle - b
|\phi_{i-1}\rangle
\label{eq: recphi}\\
b|\psi_{i+1}\rangle &=& (H-a)|\psi_{i}\rangle -
b|\psi_{i-1}\rangle + [\hat X,H]|\phi_{i}\rangle
\mbox{.}
\label{eq: recpsi}
\end{eqnarray}
%
%
Solving this latter system allows one to compute efficiently~:
%
%
\begin{equation}
A(t)|\psi\rangle=\sum_{i=0}^{\infty} c_{i} (t) |\psi_{i}\rangle
\label{eq: recP}
\mbox{,}
\end{equation}
%
%
and thus $X^2_0(E,t)$
with an arbitrary accuracy given by the order of the truncation.
%
%
%
%

\begin{figure}[h]
\centerline{\epsfxsize=100mm
\hspace{50mm}
\epsffile{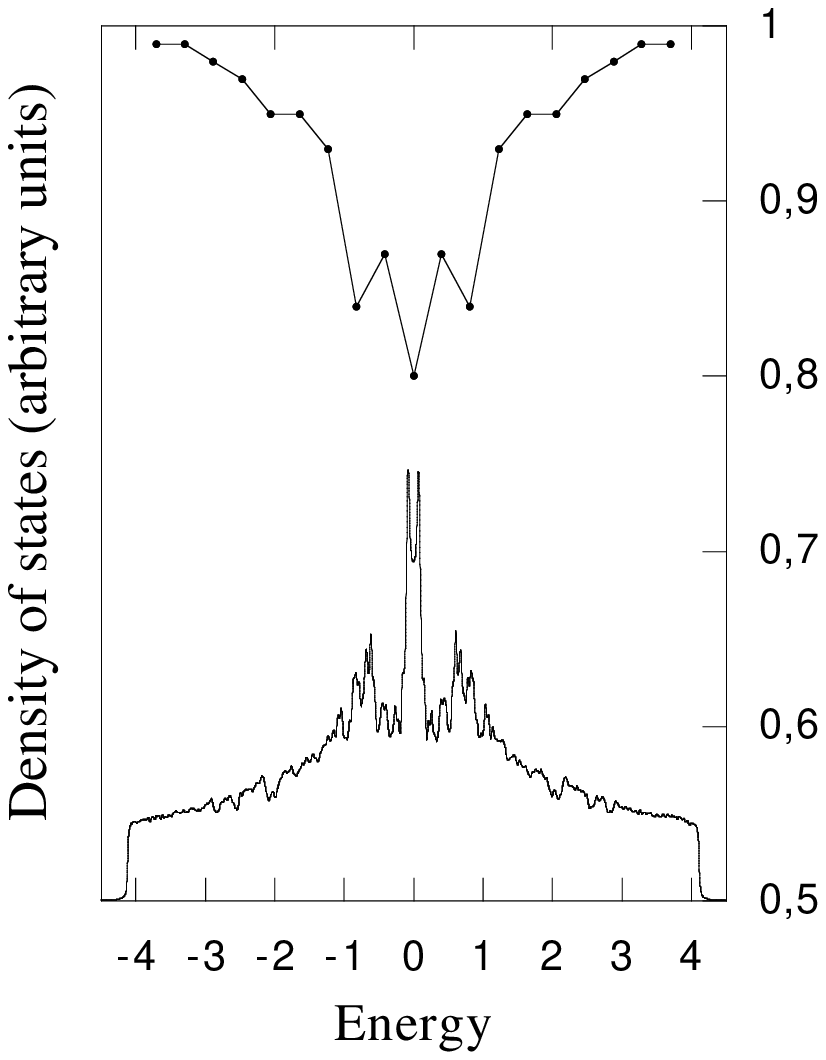}
\epsfxsize=100mm
\hspace{-176pt}
\epsffile{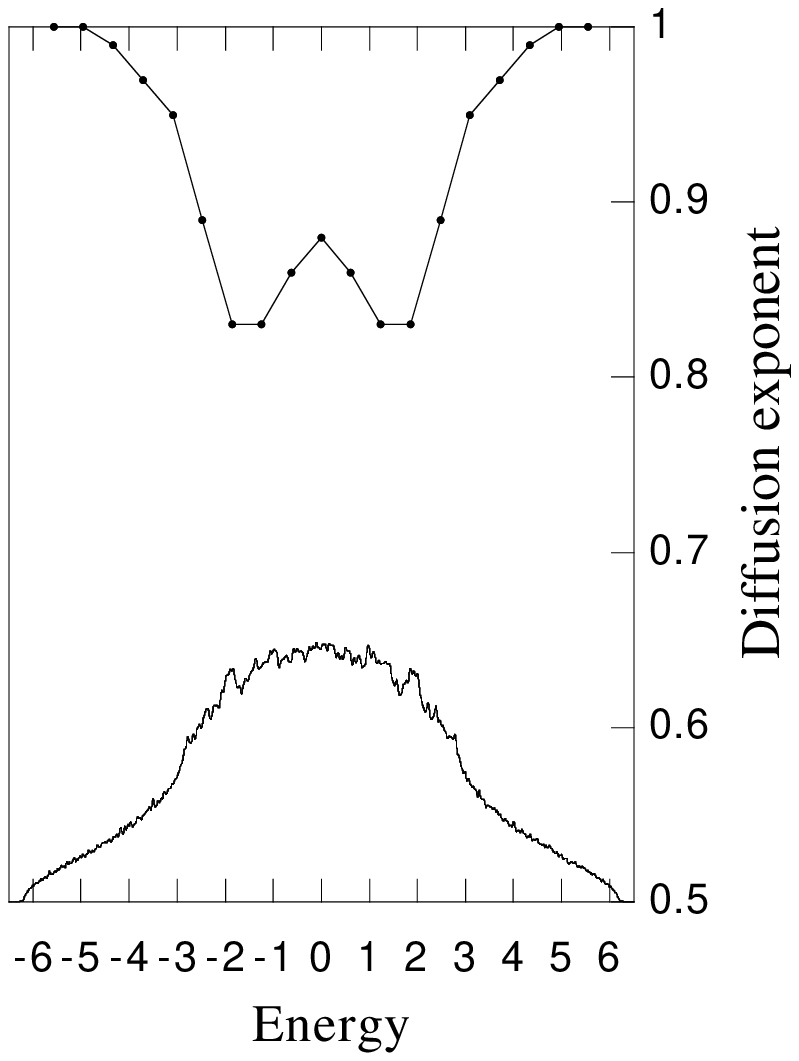}}
\vspace{-8mm}
\caption{Normalized density of states (lower curve) and diffusion exponents
(upper curve) of the $2D$ GRT (left) and of the $3D$ GRT (right).}
\label{fig:beta(E)}
\end{figure}

All the results discussed here have been obtained in the $22^{nd}$
approximant (755476 sites) for the $2D$ GRT and in the $21^{st}$
approximant ($1055026$ sites)
for the  $3D$ GRT. We have computed, for both tilings and for several
energy-filtered
wave packets, the time dependence
of $R_{0}^{2}(E,t)= X_{0}^{2}(E,t) + Y_{0}^{2}(E,t)+Z_{0}^{2}(E,t)$
where $x,y$ and $z$ are three
orthogonal directions ($Z_{0}=0$ in $2D$).
%
%
\begin{figure}
\centerline{\epsfxsize=100mm
\hspace{50mm}
\epsffile{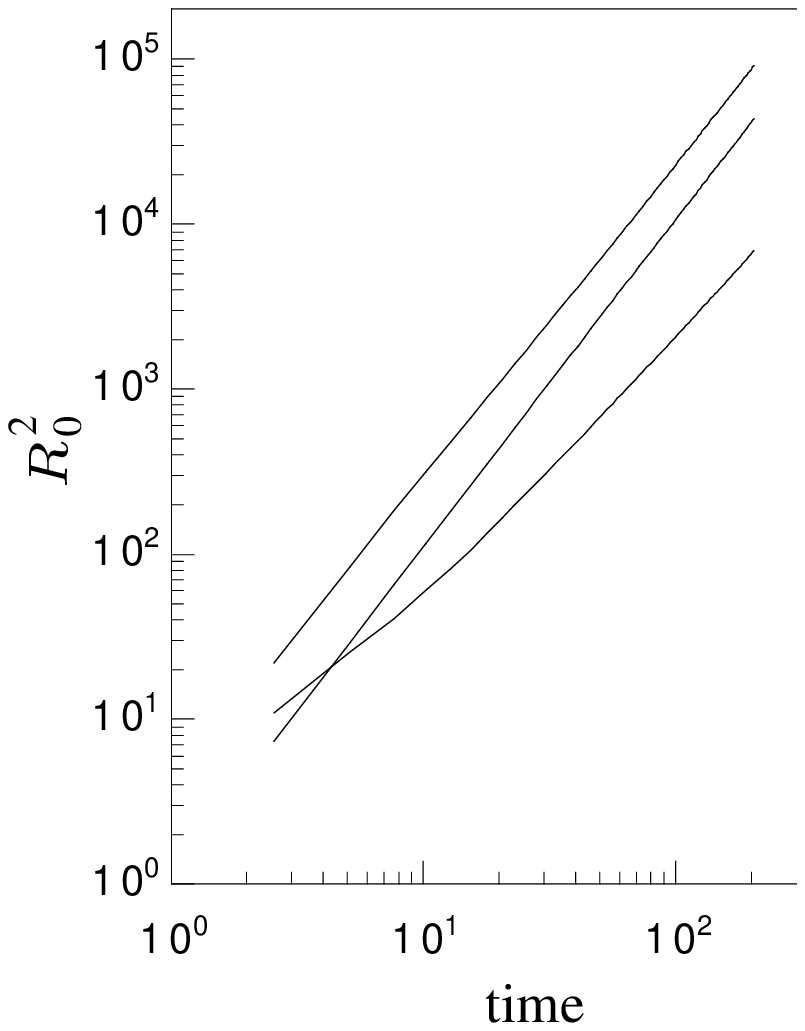}
\epsfxsize=100mm
\hspace{-176pt}
\epsffile{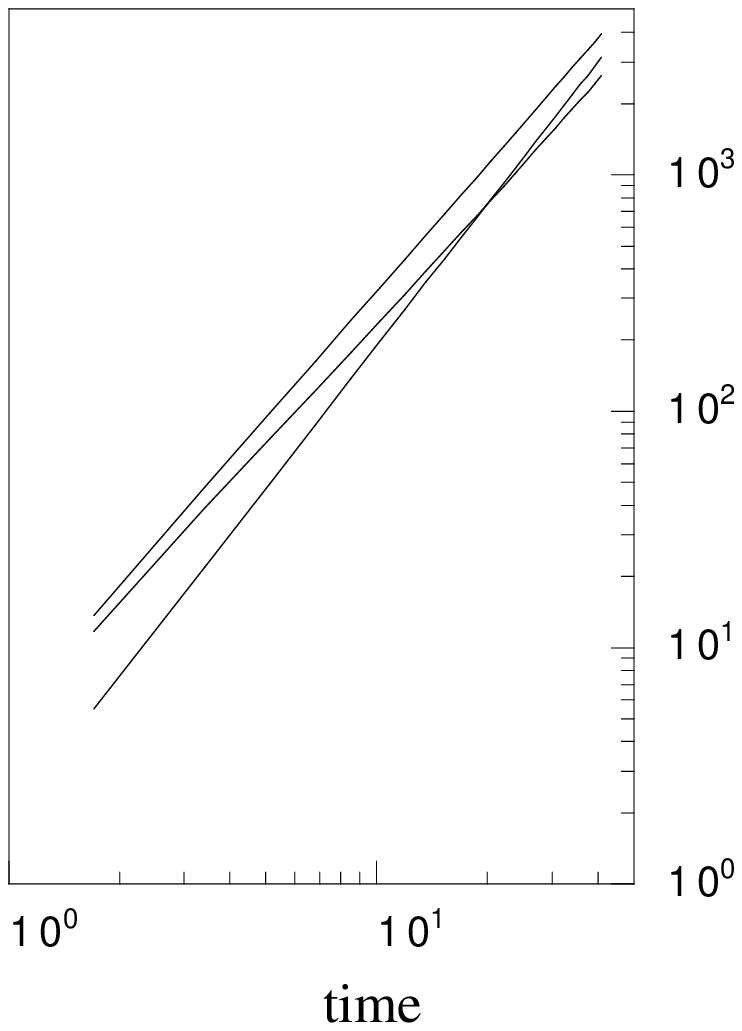}}
\vspace{-10mm}
\caption{Behaviour of $R_0^{2}(E,t)$ for three energy-filtered
wave packets in the $2D$ GRT (left) and in the $3D$ GRT (right).}
\label{fig:R2C(t)}
\end{figure}
%
%
As displayed in Fig.~\ref{fig:R2C(t)}, one has $R_{0}^{2}(E,t) \sim t^{2\beta(E)}$ at
large times with $0<\beta(E) \leq 1$. The observation of such a sub-ballistic
propagation for large approximant is consistent with the existence of critical
eigenstates.  The energy dependence of the diffusion exponent $\beta$ shown in
Fig.~\ref{fig:beta(E)} also clearly confirms the tendency observed with exact
diagonalizations since the diffusion exponent is, in both systems ($2D$ and
$3D$), larger at the band edges than at the spectrum center. In the $3D$ GRT, we
have even obtained an almost ballistic propagation $(\beta \simeq 1)$. In addition,
the shape of the density of states is nearly free- electron like at band edges which
is also consistent with the less localized eigenstates at these energies. If this
general trend is similar to what is observed in the octagonal tiling
\cite{Passaro_Octo} and in the Penrose tiling
\cite{Rieth_Penrose2D3D}, it is the opposite of what occurs in the $3D$ icosahedral
Amman-Kramer tiling
\cite{Rieth_Penrose2D3D}. Note also that in disordered systems, states at band edges
are more localized than in the band center. The parameters that determine the nature
of band edges states are obviously not yet understood.
%
%
%
%
We would like to  emphasize that the scaling laws are valid for
$R_{0}^{2}(E,t)$ varying by $3-4$ orders of magnitude in $2D$ and
$2-3$ orders of magnitude in $3D$. In both cases the number of sites
$N$ contained in a volume $R_{0}^2(E,t)$ in $2D$ and $R_{0}^3(E,t)$
in $3D$ varies
by $3-4$ orders of magnitude. This is much better than with exact
diagonalization
for which the typical variation of $N$ is $1$ order of magnitude.

The power-law behaviour obtained for $R_{0}^{2}(E,t)$ allow us to
explicitely calculate the low frequency conductivity. Indeed,
assuming $X^{2}(E_{\rm F},t)\simeq C(E_{\rm F}) t^{2\beta(E_{\rm F})}$, one obtains,
using Eq.~(\ref{eq:COND})~:
%
%
\begin{equation}
\sigma(E_{\rm F},\omega)\simeq e^{2}n(E_{\rm F}) \, C(E_{\rm F})
\,\Gamma\left(2\beta+1\right)(\mbox{i}/\omega)^{2\beta-1}
\label{eq: conduc}
\mbox{,}
\end{equation}
%
%
where $\Gamma$ is the Euler Gamma function.  For the GRT, the diffusion
exponent
$\beta(E_{\rm F})$ is always greater than $1/2$ (superdiffusive propagation) so that
although the system is non periodic, its dc conductivity
$\sigma(E_{\rm F},\omega=0)$ is infinite
for all $E_{\rm F}$.  Note that experiments are qualitatively explained if $\beta
<1/2$ \cite{Mayou00}
but such values of $\beta$ probably require a strong quasiperiodic potential.

%
%
%
%

To conclude, we would like to point that the calculation of quantum
diffusion presented
here, implies a computational load proportionnal to the number $N$ of
sites of the system
whereas exact diagonalizations implies a load which is of order
$N^{3}$ or $N^2 \ln N$ for
the best algorithms.  This is the main reason of the great efficiency
of the present
method and of its ability to treat large systems. Note that in the context of
large scale electronic structure calculations, order $N$ algorithms
raise much interest
and can also be much more efficient than exact diagonalizations
\cite{Goedecker}.
As a consequence, this approach should open new perspectives in the
investigation of $2D$
and $3D$ models.\\

We thank Claude Aslangul for a careful reading of the manuscript.


\end{document}